\documentclass[aps,twocolumn,pre,showpacs]{revtex4}
\usepackage{amssymb}
\usepackage{epsfig}
\usepackage{graphicx}
\usepackage{amssymb}
\usepackage{wasysym} 

\begin{document}

\title{Autocatalytic plume pinch-off}

\author{Michael C. Rogers$^{1}$, Abdel Zebib$^{2}$ and Stephen W. Morris$^{1}$}
\address{$^{1}$Department of Physics, University of
Toronto, 60 St. George St., Toronto, Ontario, Canada M5S 1A7\\
$^{2}$Mechanical and Aerospace Engineering, Rutgers University, Piscataway, New Jersey 09954-8058, USA}

\date{\today}

\begin{abstract}

A localized source of buoyancy flux in a non-reactive fluid medium creates a plume. The flux can be provided by either heat, a compositional difference between the fluid comprising the plume and its surroundings, or a combination of both. For autocatalytic plumes produced by the iodate-arsenous acid reaction, however, buoyancy is produced along the entire reacting interface between the plume and its surroundings. Buoyancy production at the moving interface drives fluid motion, which in turn generates flow that advects the reaction front. As a consequence of this interplay between fluid flow and chemical reaction, autocatalytic plumes exhibit a rich dynamics during their ascent through the reactant medium. One of the more interesting dynamical features is the production of an accelerating vortical plume head that in certain cases pinches-off and detaches from the upwelling conduit. After pinch-off, a new plume head forms in the conduit below, and this can lead to multiple generations of plume heads for a single plume initiation. We investigated the pinch-off process using both experimentation and simulation. Experiments were performed using various concentrations of glycerol, in which it was found that repeated pinch-off occurs exclusively in a specific concentration range. Autocatalytic plume simulations revealed that pinch-off is triggered by the appearance of accelerating flow in the plume conduit.

\end{abstract}


\maketitle

\label{intro}

Plumes are organized flow structures of considerable importance~\cite{TURNER, TURjfm} due to their widespread natural occurrence, particularly in geophysical flows~\cite{MORnat, WHIjgr, OLSjfm, GRIpfa}.  A plume can be {\it steady}, in the case when the flow structure is well-developed and is no longer time-dependent, or it can be {\it starting}, in the transient case where buoyant fluid is penetrating the medium into which it is growing. The majority of work on plumes has focused on the steady variety, since the persistence of the columnar conduit following behind the head plays a significant role in many geophysical processes. Long-lived plume conduits are believed to underlie the formation of mantle hot spots, such as the one responsible for the Hawaiian Island - Emperor Seamount chain~\cite{MORnat}.  We will consider autocatalytic starting plumes in this paper. 

Before a laminar starting plume reaches a steady state, it develops its most prominent feature: an evolving head encompassing a vortex ring. Usually, the plume head remains attached to the conduit and the conduit remains linked to the buoyancy source~\cite{GRIpfa, ROGpof}. Vortex rings are the subject of a vast literature and have a rich history~\cite{SHAarfm}. They can be produced experimentally by using a piston to transiently inject a finite volume of neutrally buoyant fluid into quiescent surroundings~\cite{GHAjfm}, a technique is similar to the vortex ring generator pioneered by P. G. Tait in the nineteenth century~\cite{TAIT}. {\it Pinch-off} refers to the process by which the vortex ring detaches from its source. In addition to neutrally buoyant scenarios, vortex ring formation and pinch-off has also been investigated in the context of buoyant starting plumes, where results suggest that dimensionless circulation of the ring is a universal quantity, regardless of whether it is produced by pinch-off from a plume head or formed by a piston~\cite{SHUjfm, POTeif}. 

Essentially free buoyant vortex rings can also be created by the pinch-off of plume heads driven by an autocatalytic chemical reaction~\cite{ROGprl}. Nonlinear chemical kinetics can produce fronts of reaction which are advected by the buoyancy-driven flow created by the reaction itself~\cite{chem_advect}.  This ``self-stirring'' can, in turn, increase the production of the product species and change the morphology of the reaction front. In a container with a relatively unconfined geometry, the localized initiation of a nonlinear autocatalytic reaction creates a special sort of reacting plume~\cite{ROGprl, ROGpre}. Such plumes can be viewed as weakly driven versions of expanding flame fronts, such as the flame ``bubbles'' that detonate Type IA supernovae~\cite{VLActm}. Unlike supernovae, autocatalytic plumes are readily produced and studied in the laboratory~\cite{ROGprl, ROGpre, MARpre}.

This paper describes experimental and numerical investigations on the behavior of autocatalytic plumes produced by the propagating front of the iodate-arsenous acid (IAA) reaction~\cite{GRBjpc, HARjpc}, focusing on the pinch-off process. The IAA reaction produces buoyancy-driven flow by its exothermicity and by the isothermal density difference between its reactant and product solutions~\cite{POJjpc}. We explore the effect of adding glycerol to the IAA solution on autocatalytic plume morphology and dynamics. Addition of glycerol increases the viscosity of the solution, and it slows the rate of diffusion of the autocatalyst. We found that pinch-off behavior is most prevalent in a specific window of glycerol concentration. Plumes produced with glycerol concentrations outside this window did not produce multiple pinch-off events. To complement the experiments, we constructed an axisymmetric numerical simulation of autocatalytic plume dynamics and used it to examine the development of the reaction-driven flow that leads to pinch-off. We conclude by identifying the subtle processes in the head and conduit that lead up to pinch-off events.

\section{Experiment}

\subsection{Apparatus}
\label{ACPexp}

\begin{figure}
\begin{center}
\includegraphics[width=6.0cm]{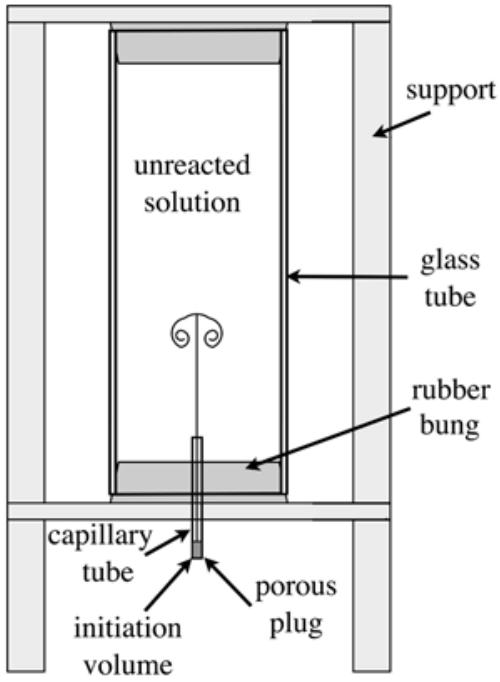}
\caption{A schematic of the apparatus. The reactant solution is contained in one of two different cylinders used for the experiments. These cylinders had lengths of 32 and 40.6~cm, and both had inner diameters of 8.9~cm. The capillary tube had an inner diameter of 2.7~mm.}
\label{apparatus}
\end{center}
\end{figure}

The apparatus used for producing IAA starting plumes is shown in Fig.~\ref{apparatus}. The reaction tank was a large glass cylinder sealed at each end by large rubber bungs. A capillary tube entered through a hole in the lower bung. The outside end of the capillary tube was sealed with a short plastic tube clamped at one end which formed the initiation volume. The plastic tube was filled by a porous plug made of loosely packed cotton. Reactions were initiated by inserting a thin hypodermic needle into the plastic tube and then injecting a very small amount of product solution into the porous plug. The purpose of the plug was to to quench any potential hydrodynamic disturbance of the quiescent reactant solution during reaction initiation. As a reaction front ascends, it must first make its way through the plug, and then into the capillary tube. Upon reaching the end of the submerged portion of the capillary tube, a reaction front is then able to escape into the large volume of reactant solution, launching a plume. The apparatus was illuminated from behind and still images of the evolving plume were captured using a digital camera. 

\subsection{Reactant preparation and fluid properties}

Reactant solutions were prepared using distilled water and reagent grade chemicals. Iodate stock solutions were made by dissolving ${\rm KIO_3}$ powder in distilled water. Arsenous acid stock solutions were prepared by dissolving ${\rm As_2O_3}$ powder in stirred water heated on a hot plate. These stock solutions were diluted so that the reactant solution contained [${\rm IO_3^-}$]=0.005M and [As(III)]=0.020M. Congo Red indicator was added to all reactant solutions in order to visualize the reaction fronts. It was present in solution at a concentration of $2 \times 10^{-5}$M. Congo Red changes color over a pH range of 3.0 to 5.0, where the acid form is blue and the base form is red. The reaction front leaves in its wake a product solution with pH of $\sim 2.7$, so that the upwelling blue products are easily visible within the red unreacted solution. In addition to the reactants, most solutions also contained chemically inert glycerol, the effect of which was to vary the fluid properties of the reactant mixtures, which are given in Table~\ref{fluprops}. Unless otherwise stated, glycerol concentrations throughout this paper are given in volume percent. Densities, $\rho$, were measured with an Anton-Paar densitometer, and kinematic viscosities, $\nu$, were determined by interpolating data from Ref.~\cite{DOW}. Values for the diffusion constant $D$ of the iodide I$^-$ autocatalyst were calculated using the Stokes-Einstein relation~\cite{einstein}, starting from the measured viscosity and the known diffusion constant of I$^-$ in water~\cite{hanna}. All experiments took place at a room temperature of $(24.0~\pm~1.0)^\circ$~C. 
\begin{table}
\centering      
\begin{tabular}{c c c c c c}  
\hline\hline                       
$\%_V$ &  $\%_M$ & $\rho$ & $\nu \times 10^{-6}$ & $D \times 10^{-9}$ & Sc $\times 10^3$\\ 
 &  & (g/cm$^3$) &(m$^2$/s) & (m$^2$/s) & \\ [0.5ex]
\hline                    
0 & 0 & 0.99745 & 0.91  & 2.0 & 0.46\\   
10 & 12.7 & 1.02787 & 1.25 & 1.4 & 0.89\\ 
20 & 24.6 & 1.05784 & 1.75 & 0.98 & 1.8\\ 
25 & 30.0 & 1.06936 &  2.07 & 0.82 & 2.5\\ 
30 & 35.8 & 1.08705 & 2.52 & 0.66 & 3.8\\ 
40 & 46.3 & 1.11483 & 3.86 & 0.42 & 9.2\\  
50 & 56.3 & 1.14046 & 6.37 & 0.25 & 25\\      
\hline     
\end{tabular} 
\caption{Fluid properties at 24.0$^\circ$C, and the Schmidt number, Sc, of the reactant fluids for various glycerol-water concentrations, given in both volume percent $\%_V$ and mass percent $\%_M$. $D$ is the diffusion constant for the iodide autocatalyst.
}
\label{fluprops}
\end{table}  

\begin{figure*}
\begin{center}
\includegraphics[height=8cm]{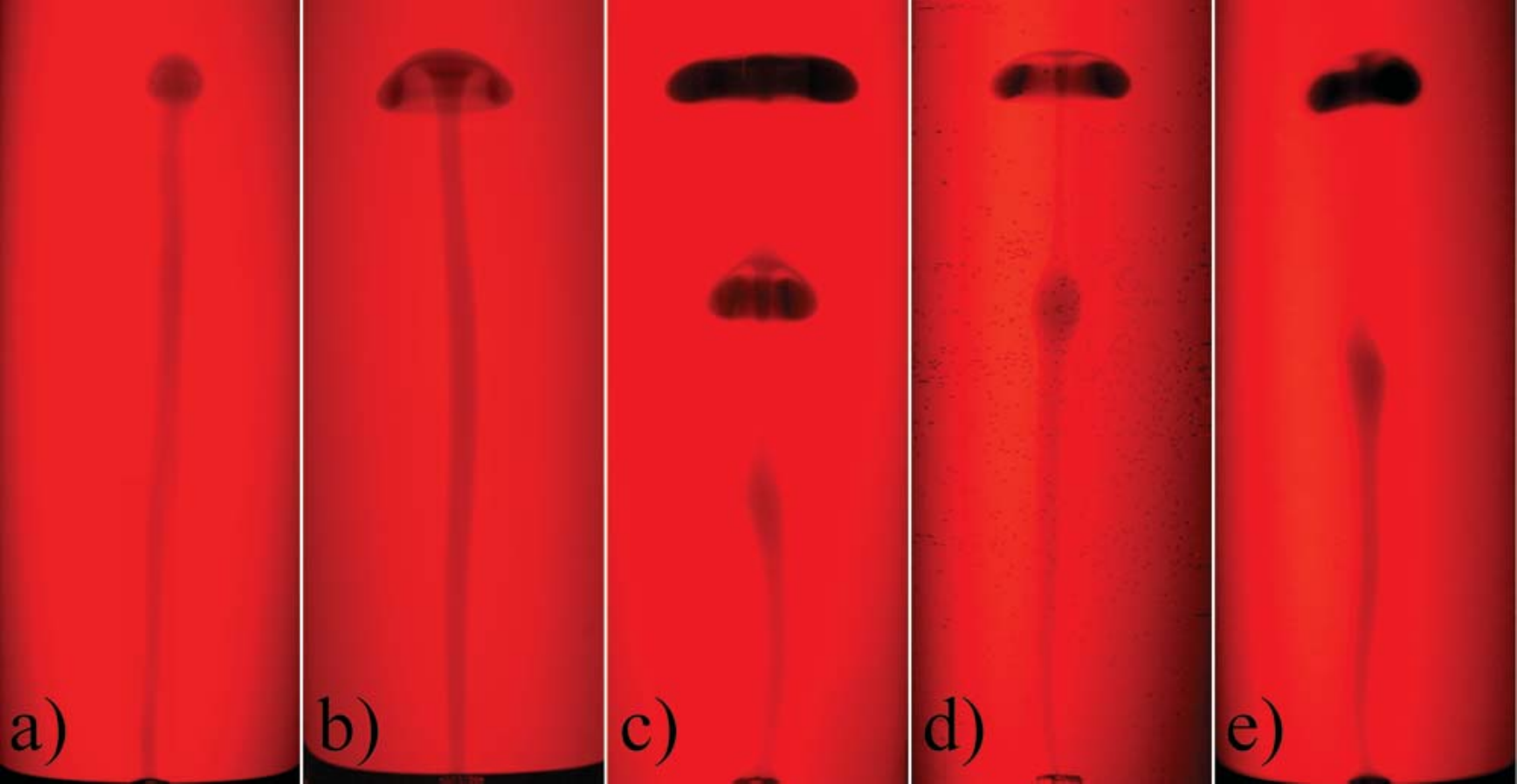}
\caption{(Color online) The morphology of autocatalytic chemical starting plumes in a) water, and in glycerol concentrations of b) 20\%, c) 30\%, d) 40\%, and e) 50\%. All images are of plumes at a height of $\sim$20 cm from the outlet. It took the plumes a) 255~s  , b) 555~s, c) 425~s, d) 449~s, and e) 681~s to ascend to this height before these images were captured.}
\label{morph}
\end{center}
\end{figure*}

\subsection{Starting plume morphology and pinch-off}

After escaping the capillary tube, the advancing autocatalytic reaction front generates a starting plume that accelerates upward and may exhibit one or more pinch-off events, depending on the experimental conditions.  The various stages in the evolution of a plume that pinches-off were first described in Ref.~\cite{ROGprl}. In summary, these stages are as follows. In the initial stage the plume creeps slowly out of the capillary tube and its head remains roughly spherical. Following this stage, the plume head changes shape and the vortical motion within the head becomes visible as the product solution wraps within an overturning vortex ring. As the plume head continues to ascend and enlarge, it begins to pinch-off from the upwelling conduit. During this stage of growth, the head becomes almost completely detached from the conduit, and becomes an essentially free vortex ring. Beneath this vortex ring, a rising bulge in the conduit forms which eventually becomes a second generation plume head. Multiple generations of plume heads can be formed through this pinch-off process.

Whether or not an autocatalytic plume head pinches-off, and also the number of pinch-offs, depends on the fluid properties of the reactant solution. Fig.~\ref{morph} shows the morphologies of fully developed starting plumes produced in solutions with various concentrations of glycerol. Plumes produced in the solutions without glycerol, such as the one shown in Fig.~\ref{morph}a, did not exhibit pinch-off. Similarly, in 20\% glycerol solutions pinch-off was not observed. All plumes that did exhibit pinch-off, such as those shown in Fig.~\ref{morph}c-e, generally had similar morphologies, regardless of the number of times that pinch-off occurred. For experiments in 30\% and 40\% glycerol, there were always two or more pinch-offs. At glycerol concentrations of 50\%, however, pinch-off was rarely observed. In cases where it did happen, such as the run shown in Fig.~\ref{morph}e, only a single pinch-off event took place. 

\begin{figure}
\begin{center}
\includegraphics[height=6.5cm]{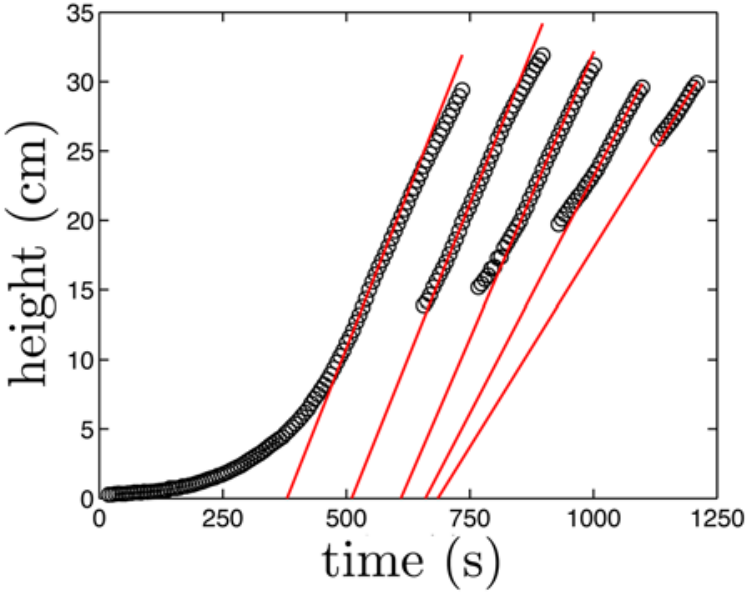}
\caption{(Color online) The height as a function of time for five plume heads produced by a 40\% glycerol run in the 40.6 cm long cylinder. Lines indicate linear fits to the five consecutive data points with the maximum slope. Slopes of lines such as these were used to determine $v_{max}$ for each plume head.}
\label{five_heads}
\end{center}
\end{figure}

In addition to fluid properties, the number of pinch-offs observed for a given reactant solution composition also depended on the length of the cylinder in which experiments were performed. In the 32 cm high cylinder, the maximum number of fully developed plume heads produced was four. This occurred for a run in a 30\% glycerol solution. To test whether or not the number of pinch-offs could be increased by giving plumes more vertical space to develop, experimental runs were performed in an apparatus with the same diameter as the 32 cm long cylinder, but with a length of 40.6 cm. These runs were only performed for glycerol concentrations of 20\%, 40\%, and 50\%. For 50\% glycerol plumes, no additional plume heads were generated. However, for 20\% glycerol plumes, extending the length allowed for a single pinch-off to occur, when, in contrast, no pinch-offs were observed in the shorter cylinder for the same solution. The maximum number of fully developed heads for a 40\% glycerol run in the longer cylinder was five, compared to just three heads in the shorter cylinder. The ascent data for five plume heads produced in a 40\% glycerol solution is shown in Fig.~\ref{five_heads}. 

\subsection{Plume head Reynolds number}

For each experimental run, the ascent velocity of each plume head was calculated using linear fits to the head height as a function of time.  The maximum head velocity $v_{max}$ was determined from the fit to the five consecutive data points giving the largest slope. From $v_{max}$, the head Reynolds number, 
\begin{equation}
{\rm Re}_h=v_{max}w_h/\nu
\end{equation}
was calculated, where $w_h$ is the width of the plume head at the average time over which $v_{max}$ was measured. $w_h$ was measured from digital images, taking into account the optical magnification due to the cylindrical shape of the reaction tank. We observed that Re$_h$ had similar values for all of the plume heads produced in a given run. The exception to this generality was the value of Re$_h$ for the final head produced in a run with multiple heads, which was always somewhat less than the others. This was mainly due to the smaller value of its $v_{max}$. Typically, for a given experimental run, the final plume head would develop close to the top of the tank, giving it less time to accelerate before it hit the top surface of the fluid. This limitation is enhanced by the growing mass of product solution at the top of the tank which further decreases the vertical space into which the final head can ascend. For all of our experiments, Re$_h$ never exceeded $\sim 20$. This value was reached in a 20\% glycerol run, where $w_h \sim 2.9$~cm, $v_{max} \sim 0.12$~cm/s, and the kinematic viscosity of the solution was $\nu = 1.75 \times 10^{-2}$~cm$^2$/s. The value of Re$_h$ indicates that while the flow around the plume head is certainly laminar, Re$_h$ is not so small that it can be considered Stokes or ``creeping" flow, which would require Re$_h \ll 1$.

To understand the general tendency of the plumes to pinch-off, we examined the trend in the number of pinch-offs and head Reynolds numbers for varying glycerol concentration. The glycerol concentration is conveniently expressed in terms of the dimensionless Schmidt number, 
\begin{equation}
{\rm Sc}=\nu/D. 
\label{schmidt}
\end{equation}
Recall that the glycerol concentration effects both the kinematic viscosity $\nu$ and the molecular diffusivity of the iodide autocatalyst $D$.  The complete dynamical equations and their nondimensionalization are discussed in Section \ref{scaling} below.

\begin{figure}
\begin{center}
\includegraphics[width=8.5cm]{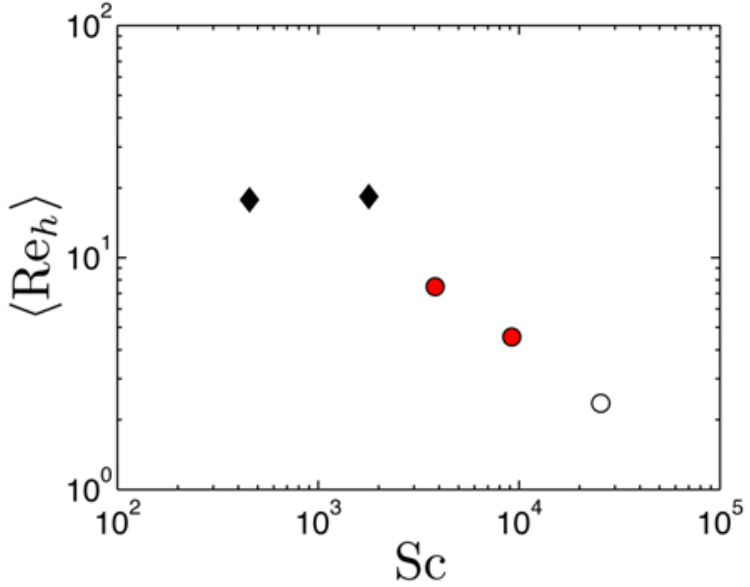}
\caption{(Color online) The average plume head Reynolds number $\langle{\rm Re}_{h}\rangle$ for glycerol solutions characterized by various  Schmidt numbers, Sc. Diamond symbols indicate runs that did not pinch-off in the short cylinder apparatus. Circular points indicate that pinch-off occurred, with shaded (red) points indicating conditions under which two or more pinch-offs were observed.}
\label{resc}
\end{center}
\end{figure}

For all experimental runs, the mean of the plume head Reynolds number $\overline{{\rm Re}}_h$ for each run was calculated by averaging the Re$_h$ values for all the heads, omitting the final one in the case of multiple pinch-offs. As discussed above, the last head often did not have the time to fully develop before interacting with the top of the tank. $\overline{{\rm Re}}_h$ values from different runs were then averaged to obtain a characteristic Reynolds number $\langle{\rm Re}_{h}\rangle$ for all experiments that took place in fluids with same glycerol concentration and hence the same value of Sc. These data are shown in Fig.~\ref{resc}. Examples of the plume morphologies for each of the Schmidt numbers displayed in Fig.~\ref{resc} are shown in Fig.~\ref{morph}. 

For the two smallest Schmidt numbers, corresponding to runs in 0\% and 20\% glycerol, $\langle{\rm Re}_{h}\rangle$ is approximately independent of Sc and no pinch-offs were observed. This reflects the rather small and undeveloped head morphology under these conditions. As Fig.~\ref{morph}a shows, in pure water, the head remains small and compact. In 20\% glycerol, a vortex ring develops in the head only near the top of the tank. While these plumes rise faster than plumes in solutions containing more glycerol, $\langle{\rm Re}_{h}\rangle$ is limited by the small size of $w_h$. As glycerol concentration increases, Sc increases and $\langle{\rm Re}_{h}\rangle$ begins to drop. In this regime, multiple pinch-offs are observed. This behavior is accompanied by a decrease in $\langle{\rm Re}_{h}\rangle$ with viscosity with only small changes in $w_h$ and $v_{max}$. Finally, at the highest value of Sc, only one pinch-off is observed while $\langle{\rm Re}_{h}\rangle$ continues its downward trend. Thus, we find that multiple pinch-off behavior occurs in a special, intermediate range of Sc. 

Plume heads reach maximum velocity because as the head grows in width, the flow field begins to have stronger viscous interactions with the cylindrical wall of the tank. As the head reaches widths comparable to the diameter of the cylinder, vortical fluid motion in the head appears to quench because of wall effects. Eventually, the constrained vortical motion is no longer able to effectively entrain fresh reactant solution into the head, slowing the growth of buoyancy generated by the reaction. In the intermediate range of Sc, the plume rises slowly enough that a complex head morphology and multiple pinch-off events have time to develop. For the highest value of Sc, however, the increased constraint from boundary effects is likely the reason why pinch-off is seldom observed. 

\begin{figure}
\begin{center}
\includegraphics[height=6.5cm]{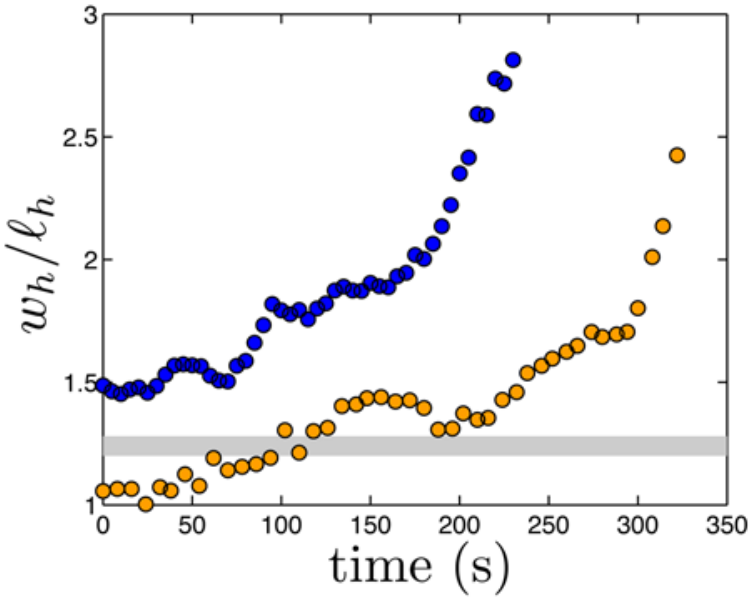}
\caption{(Color online) The aspect ratio $w_h/\ell_h$ for autocatalytic plumes in a 30\% (dark circles, blue online) and 40\% (light circles, orange online) glycerol solution. The horizontal grey line indicates the constant value of $w_h/\ell_h=1.24 \pm 0.04$ for non-reacting forced plumes~\cite{ROGpof}. Here, $t=0$ corresponds approximately to when a vortex ring first forms in the head, not to the moment when the reaction front emerges from the capillary tube.}
\label{wlchem}
\end{center}
\end{figure}

\subsection{Evolution of head shape}

In addition to the head width $w_h$, the head length $\ell_h$, was also measured from digital images. We defined $\ell_h$ as the distance from the top of the head to the base of the underturning lobe beneath it. The dimensionless aspect ratio $w_h/\ell_h$ is a useful way to compare autocatalytic plume heads to buoyant plume heads produced without chemical reactions~\cite{ROGpof}. For laminar forced plumes, the growing plume heads are self-similar and this ratio is constant~\cite{ROGpof} with $w_h/\ell_h=1.24 \pm 0.04$. In contrast, autocatalytic plume heads do not remain self-similar.  Fig.~\ref{wlchem} shows that the aspect ratio of the autocatalytic plume head increases with time. A steep increase in the rate of change of $w_h/\ell_h$ occurs at approximately the same time that a narrowing of the conduit beneath the head becomes obvious, indicating that a pinch-off event is under way.

\begin{figure}
\begin{center}
\includegraphics[height=6.5cm]{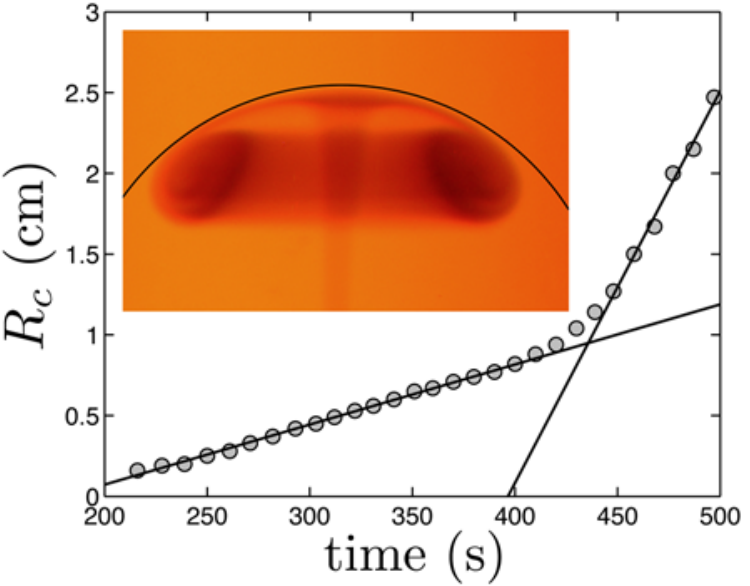}
\caption{(Color online) The radius of curvature $R_c$ of the crest of an autocatalytic plume head as a function of time. The two straight lines are fits to the data before and after pinch-off. The inset shows the crest of an autocatalytic plume head and the best fit circle that describes its shape. The head was imaged at a height of 16.7 cm above the outlet, 448 s after the front emerged from the capillary.}
\label{curve_time}
\end{center}
\end{figure}

Another interesting feature of autocatalytic plume head morphology is the spherical shape of the top of the head, which we refer to as the {\it crest}. The inset of Fig.~\ref{curve_time} shows an image of a 40\% glycerol plume head with a circle fitted to the crest shape. Positions extracted from the image using edge detection were corrected for the optical distortion of the cylindrical tank and then fit to a circle to extract the radius of curvature $R_c$. This analysis is similar to the method employed in a classic study of the shape of bubbles rising in a cylinder by Davies and Taylor~\cite{DAVprs}. The growth of $R_c$ with time for a 40\% glycerol plume is shown in Fig.~\ref{curve_time}. We find that $R_c$ goes through two distinct periods of linear growth. This is in contrast to the growth of the head width $w_h$, which has been shown in previous work to have three distinct growth regimes~\cite{ROGprl}. While the transition from a small head without vortical motion to one with a vortex ring marks a change in growth rate of the width, it does not cause a change in the growth rate of the crest curvature $R_c$. Instead, the only transition observed in the growth rate of $R_c$ occurs after the head pinches-off from the conduit and becomes an almost-free vortex ring. A steep increase in growth rate of $R_c$ is a very clear and unambiguous morphological indicator of pinch-off.

\section{Autocatalytic Plume Simulation}

In this section, we turn to a discussion of the axisymmetric numerical simulation which was used to diagnose the processes leading up to pinch-off in autocatalytic plumes. While it was computationally intractable to use our numerical model to simulate fully realistic plumes with multiple pinch-off events, it was possible to use our model to gain insight into the main features of the process.

\subsection{Dimensionless scaling considerations}
\label{scaling}

To describe our numerical model, it is helpful to first focus on the various dimensionless groups which control the fluid mechanical, thermal, and chemical aspects of plume dynamics. Plume motion depends on three different diffusivities. The most rapidly diffusing quantity is the momentum, which scales with the kinematic viscosity $\nu$. The reaction front is exothermic, and the diffusion of heat scales with the thermal diffusivity $\kappa$. Finally, the concentration of the autocatalyst is the slowest diffusing quantity, scaling with the diffusivity $D$.

In addition to the Schmidt number ${\rm Sc}$, defined in Eqn.~\ref{schmidt}, a second independent dimensionless ratio is the Lewis number,
\begin{equation}
{\rm Le} = {\kappa}/{D}. 
\label{lewis}
\end{equation}
The Prandtl number, given by
\begin{equation}
{\rm Pr} = {\rm Sc}/{\rm Le} =  \nu/\kappa, 
\label{prandtl}
\end{equation}
is not required in our formulation, but naturally appears in discussions of purely thermal plumes~\cite{TURNER}.

For a 40\% glycerol solution, using $\kappa = 1.2 \times 10^{-3}$~cm$^2$/s and values from Table~\ref{fluprops}, we find
\begin{equation}
{\rm Sc} = 9000 { , } ~~{\rm Le} = 280 {,} ~~{\rm and} ~~{\rm Pr} = 32. 
\label{lewis}
\end{equation}
The large values of ${\rm Sc}$ and ${\rm Le}$ indicate that momentum and heat diffuse much faster than chemical concentration, and hence the concentration profile at the reaction front is extremely sharp compared to all the other fields that surround an autocatalytic plume. In general, the wide range of Sc, Le and Pr indicate strong scale separation between the various effects.

The IAA reaction produces buoyancy by means of both temperature and concentration changes. These are both sufficiently small that we may employ the Boussinesq approximation,  with the density of solution
\begin{equation}
\rho=\rho_0[1-\alpha_T (T_1-T_0) - \alpha_c (c_1-c_0)],
\label{boussinesq}
\end{equation}
where $\rho_0$ is the initial density at reference temperature $T_0$ and product concentration $c_1=c_0$, with $c_0$ being the initial concentration of reaction product in the solution. $\alpha_T$ and $\alpha_c$ are the thermal and compositional expansion coefficients, respectively. In this approximation, Eqn.~\ref{boussinesq} implies that the density is a linear function of the temperature and product concentration of the solution.

Across the reaction front, there are both temperature and concentration changes, which we denote by $\Delta T = T_1-T_0$ and $\Delta c = c_1 - c_0$, respectively. In order to quantify the thermal and compositional contributions to buoyancy separately, we define two Rayleigh-like numbers~\cite{DHEprl, DHEjfm}. The thermal and concentration Rayleigh numbers are 
\begin{eqnarray}
{\rm Ra}_T &=& \frac{g \alpha_T L^3}{\nu D} \Delta T ~~{\rm and}~~~{\rm Ra}_c = \frac{g \alpha_c L^3}{\nu D} \Delta c,
\end{eqnarray} 
respectively.  Here, $L$ is a length scale that we choose to be equal to the diffusive thickness $\ell$ of the reaction front, defined by $\ell=\sqrt{D\tau}$, where $\tau$ is a reaction time scale. We calculated $\ell$ by measuring the front velocity, $v_f$, in a capillary tube where convective effects were absent, and by defining $v_f=\sqrt{D/\tau}$ and $\tau=\ell/v_f$. Parameters for the simulation were for a 40\% glycerol solution, in which we measured $v_f=9.9 \times 10^{-4}$~cm/s. Using $D= 4.2 \times 10^{-10}$~m$^2$/s from Table~\ref{fluprops},  we found $\tau= 4.3$~s and $\ell= 4.2 \times 10^{-3}$~cm. These values, along with $\alpha_T=4.3 \times 10^{-4}$ K$^{-1}$ and $\Delta T = 0.5$~K~, estimated from values given in Refs.~\cite{DOW, EDWpra}, give Ra$_T=0.10$. We measured $\alpha_c \Delta c = 3.6 \times 10^{-4}$, which along with the other relevant parameters gives Ra$_c=0.17$. Given the similarity between the values of Ra$_T$ and Ra$_c$, we see that the buoyancy effects from temperature and concentration changes are comparable in our system. 

\subsection{Numerical model}

\begin{figure}
\begin{center}
\includegraphics[height=6cm]{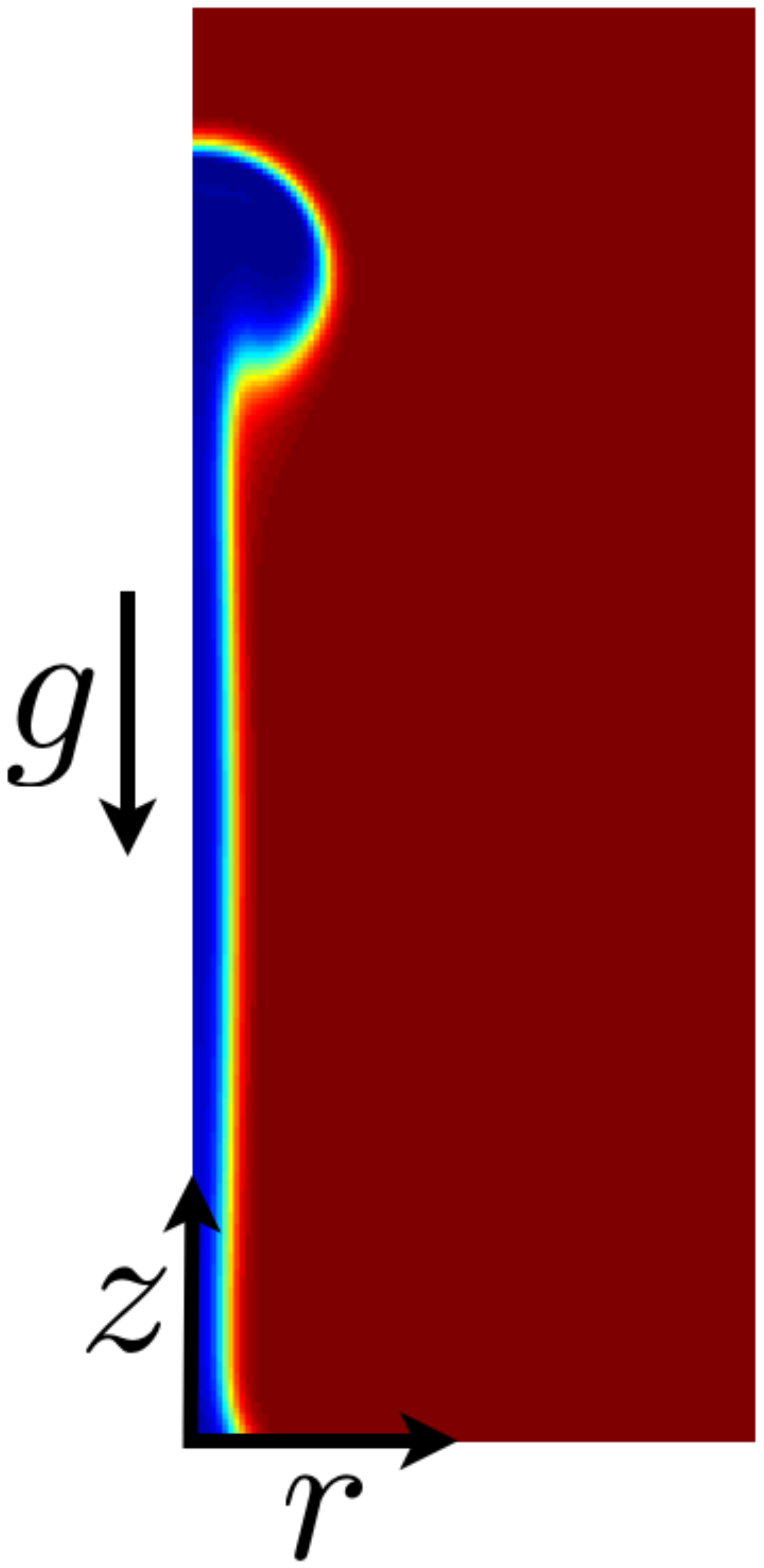}
\caption{(Color online) The cylindrical coordinate system used, with the concentration field for a plume produced by the simulation.}
\label{cyl_coords}
\end{center}
\end{figure}

Our simulation of autocatalytic plume dynamics was based on a well-established model describing flow driven by an autocatalytic reaction in a thin vertical slot~\cite{DHEprl, DHEjfm}. The autocatalytic plume simulation was axisymmetric about the direction of gravity and carried out in the cylindrical coordinates $(r,z)$ shown in Fig.~\ref{cyl_coords}, which also shows an example of the concentration field of a numerical plume.
The full dimensionless equations for our autocatalytic plume model are
\begin{equation}
\mathbf{\nabla} \cdot \mathbf{u} = 0,
\label{continuity}
\end{equation}
\begin{equation}
\frac{1}{\rm {Sc}} \frac{D\mathbf{u}} {Dt} = - \mathbf{\nabla} p + \mathbf{\nabla} ^2 \mathbf{u} + (Ra_{T}T+Ra_{c}c)\mathbf{e}_{z},
\label{NSE}
\end{equation}
\begin{equation}
\frac{Dc} {Dt} = \mathbf{\nabla}^2c + F(c),
\label{cEQN}
\end{equation}
and
\begin{equation}
\frac{DT} {Dt} = {\rm Le}\mathbf{\nabla}^2T + F(c).
\label{TEQN}
\end{equation}
Here, $\mathbf{e}_{z}$ is an upward pointing unit vector and $D/Dt$ is the material derivative.
Equations~\ref{cEQN} and \ref{TEQN} incorporate  a nonlinear reaction term given by
\begin{equation}
F(c) = c^2(1-c),
\end{equation} 
which describes the cubic autocatalytic IAA reaction. In this equation, the concentration $c$ is dimensionless and comes from scaling $c_1$, the concentration of the autocatalytic product of the reaction, by its maximum value $c_{max}$. The fully reacted fluid therefore has a dimensionless product concentration of $c = c_1/c_{max}=1$, and completely unreacted fluid has $c=0$. The dimensionless temperature from the reaction $T$ in eqs.~\ref{NSE} and \ref{TEQN} is bounded such that $0 \le T \le 1$, and comes from scaling $\Delta T$ by $\Delta T_H = - \Delta H \Delta C/\rho_0 C_p$. Here, $\Delta H$ is the heat of reaction and is negative for the exothermic reactions, and $C_p$ is the specific heat at constant pressure. 

The cylindrical computational domain was bounded by $r=r_b$,  $z=0$ and $z=z_b$. Non-slip boundary conditions were imposed on the velocity field, and no flow conditions on the energy and chemical concentrations. These conditions are given by 
\begin{eqnarray}
{\bf u}&=&\frac{\partial T}{\partial r}=\frac{\partial c}{\partial r}=0~~~{\rm at}~~~r=r_b,\\
{\bf u}&=&\frac{\partial T}{\partial z}=\frac{\partial c}{\partial z}=0~~~{\rm at}~~~z=0~~{\rm and}~~ z=z_b.
\end{eqnarray}

The numerical calculation used the primitive velocity variables $\mathbf{u} = (u,v)$ and the non-hydrostatic pressure $p$, and was carried out using the ``semi-implicit method for pressure linked equations" (SIMPLE) algorithm~\cite{thesis_ref_107, thesis_ref_108}. 
\begin{figure}
\begin{center}
\includegraphics[width=8.5cm]{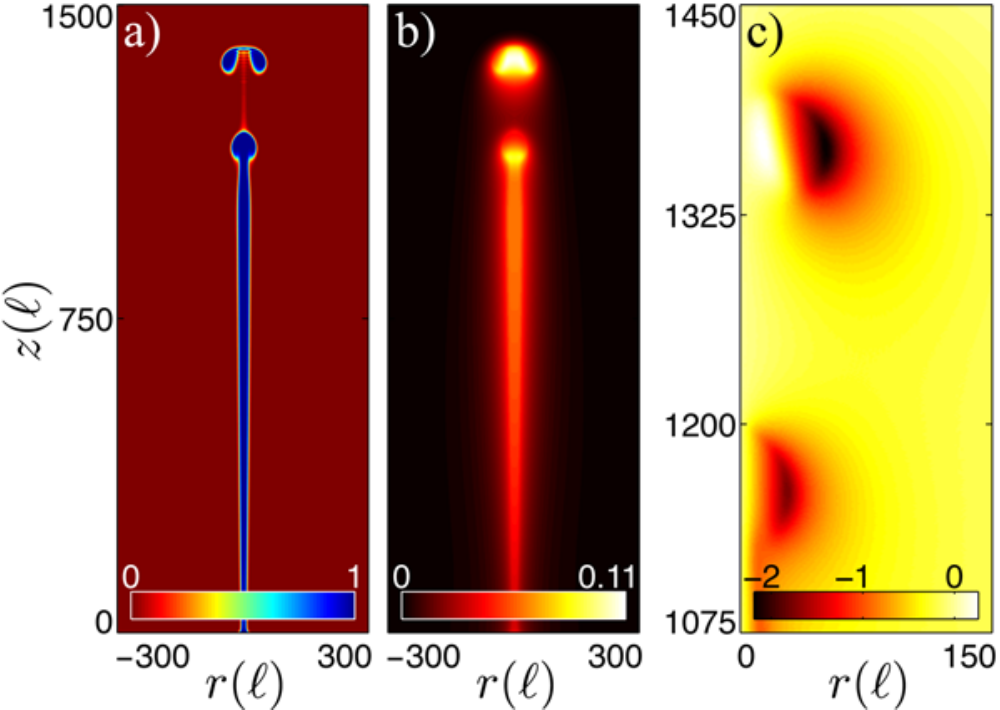}
\caption{(Color online) The a) concentration, b) temperature, and c) vorticity fields for an autocatalytic plume that has pinched-off. The concentration profile shows a plume with $c=1$ separated from the surrounding fluid with $c=0$ by a thin reaction front. The spatial domain for the simulation was $r_b=500\ell$ and $z_b=2000\ell$, only data in the region bounded by 
$r=300\ell$ and $z=1500$ is shown. The time after initiation is $42.5\tau$.}
\label{all_fields}
\end{center}
\end{figure}

\subsection{Simulation results}

In this section, we compare our simulation results to the phenomenology of 40\% glycerol plumes discussed previously. We observed the growth and pinch-off of computational plumes, and all aspects of their dynamics and morphology were in good, albeit qualitative, agreement with experimental plumes. The main limitation of the numerical computation is in the details of the initial conditions and in the size of the computational domain, which was necessarily smaller than that of the experiment. An example of the concentration, temperature, and vorticity fields from the simulation of an autocatalytic starting plume are shown in Fig.~\ref{all_fields}. 

The exact details of the inlet pipe used to launch plumes experimentally were not simulated. Instead, the initial conditions used for the simulation were a spherical region of product with initial radius $r=15\ell$ within which $c = T = 1$ at $t=0$. The spherical region was placed just above the bottom boundary so that its centre was initially located at $z=16\ell$ and $r=0$. The full spatial domain used for the simulation had $r_b=500\ell$ and $z_b=2000\ell$, and grid points with spacings of $\ell$. The time step was $\Delta t=0.005\tau$. This computational domain corresponds to a physical cylinder with a diameter of 4.2~cm and a height of 8.4~cm. These dimensions correspond to approximately one half of the diameter and one quarter of the height of the experimental apparatus. The computational expense was too great to extend the spatial domain beyond these dimensions. Although the region of the reacted fluid remained far from the boundaries, as it was for the experimental plumes, the region of nonzero fluid velocity naturally extended to the boundaries and thus we may expect that the computational plume flow was more confined than in the experiments.

The concentration field in Fig.~\ref{all_fields}a shows a plume with a similar morphology to some of the experimental plumes shown in Fig.~\ref{morph}. At this stage of the evolution of the plume, it has already undergone the pinch-off process and a new, second generation head has started to form. The temperature field around the plume head in Fig.~\ref{all_fields}b is much more diffuse than the concentration field comprising the head. This reflects the large value of the Lewis number, which is the ratio of these two diffusivities. Interestingly, the warmest region of fluid is only partially enclosed by the reaction front, while most of it resides in the center of the vortex ring. This effect is similar to the localization of hotspots in the cusps of two dimensional fingered fronts~\cite{grosfils}. This central location also happens to be where the only positive vorticity appears. The vorticity field for a plume that has pinched-off is shown in Fig.~\ref{all_fields}c. There are two regions with negative minima in the vorticity field, located in the cores of the vortex rings, where there is strong clockwise fluid motion. The largest negative vorticity is in the vortex core of the first generation head and the second minimum is located where the vortex ring in the second generation head is forming. 

\begin{figure}
\begin{center}
\includegraphics[width=8.2cm]{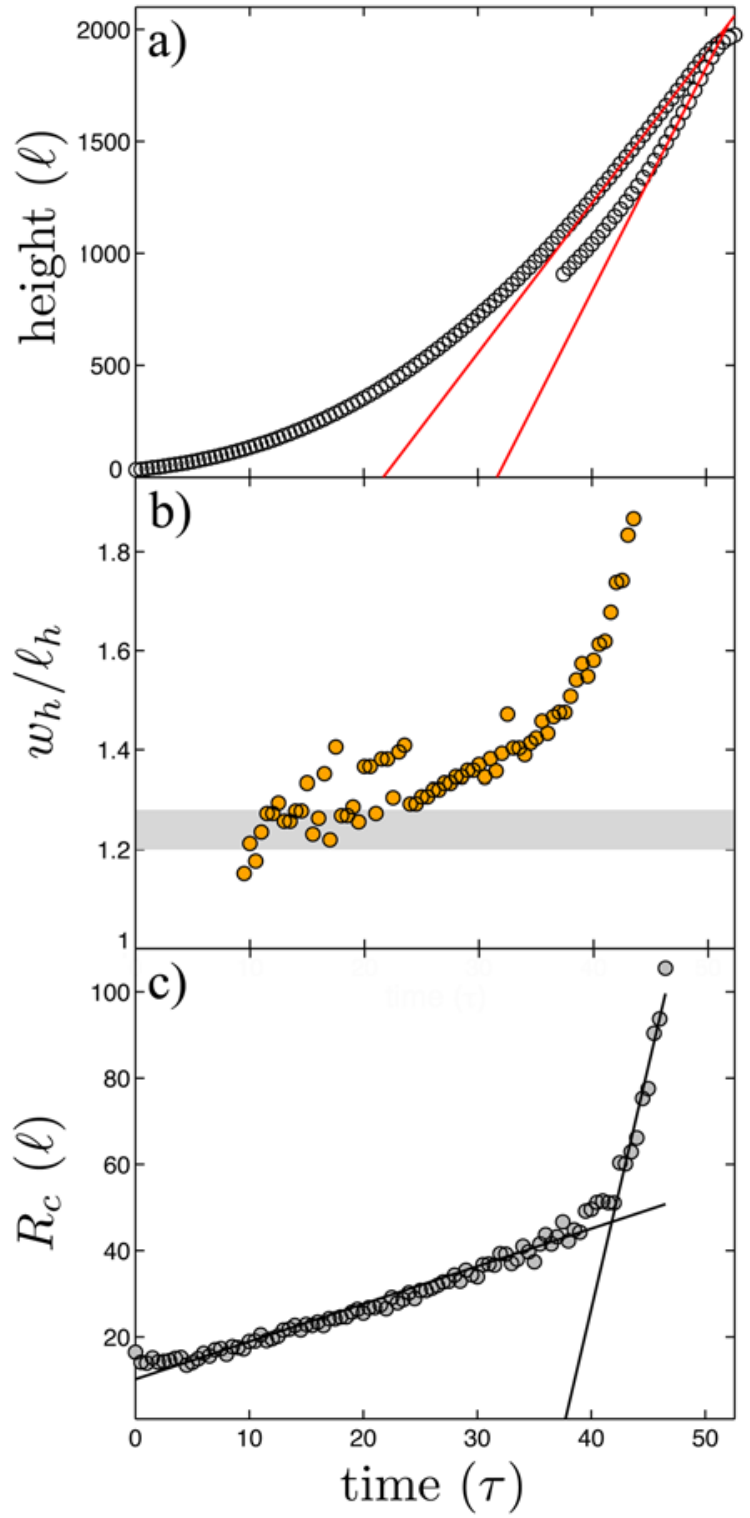}
\caption{(Color online) a) The height of first and second generation simulated plume heads as a function of time.  b) The width to length $w_h/\ell_h$ aspect ratio and c) the radius of curvature $R_c$ of the crest of the first generation head. The shaded region in (c) shows the constant aspect ratio of non-reacting forced plumes~\cite{ROGpof}. Parts a), b), and c) show simulation data that correspond to the experimental data presented in Figs. \ref{five_heads}, \ref{wlchem}, and \ref{curve_time}, respectively.}
\label{sim_stack}
\end{center}
\end{figure}

Figure~\ref{sim_stack} shows that there is good qualitative agreement in the general behavior of experimental and simulated autocatalytic plume heads. Fig.~\ref{sim_stack}a shows that the simulated plume head accelerates as it ascends, even after it has pinched-off from the conduit, which was also observed experimentally. 
The pinched-off head continues to accelerate until it reaches a maximum velocity, after which time it decelerates because of its interaction with the boundary at the top of the computational domain. The maximum velocity that the first generation head achieved in this simulation was $v_{max}=67\ell/\tau=0.065$~cm/s. This compares reasonably well with $v_{max}=0.090$~cm/s extracted from the experimental data for the first generation plume head in Fig.~\ref{five_heads}d. It is not surprising that the value of $v_{max}$ achieved by the simulated plume is less than that of the experimental plume because of the difference in boundary constraints between the two. The simulated plume has greater viscous interaction with the boundaries, and it also has less vertical space to grow and accelerate.   

The simulation also shows good qualitative agreement between the morphological characteristics of simulated and experimental autocatalytic plume heads. Simulation results for $w_h/\ell_h$ and $R_c$ are shown in Fig.~\ref{sim_stack}b-c, which should be compared to Figs.~\ref{wlchem} and~\ref{curve_time}~respectively. 

\subsection{Simulated pinch-off}

\begin{figure}
\begin{center}
\includegraphics[width=9cm]{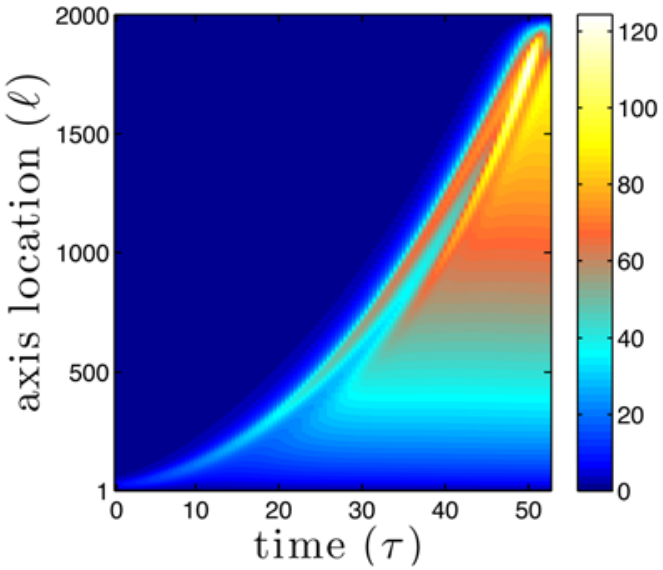}
\caption{(Color online) A space-time diagram showing the development of the centerline velocity $v_c$ of a simulated plume. The units of $v_c$ are $\ell/\tau$.}
\label{spacetime}
\end{center}
\end{figure}

In their early stages, the external appearances of autocatalytic plumes are similar to non-reacting, laminar plumes~\cite{TURNER, ROGpof, GRIpfa}.
This suggests that the interior flow profiles of autocatalytic and non-reacting laminar plumes might be similar, at least initially. The simulation results, however, show that this is not the case. A non-reacting laminar plume head is formed by buoyancy flux from its upwelling plume conduit. 
In order to feed the head, the conduit flow must rise at a greater velocity than the head~\cite{ROGpof}. In this scenario, the plume head cannot detach from the conduit. As the head rises, it must force its way up through quiescent fluid, and this induces large scale return flow in the tank. Once this flow has been established below the plume head, fluid rising in the conduit experiences less resistance to motion than was initially experienced by the head --- the conduit is moving upwards into fluid that already has momentum in the direction of its motion. This leads to the formation of a steady-state conduit flow below the plume head, where the velocity profile across a horizontal cross section of the conduit is constant, independent of height~\cite{ROGpre}. 

The simulation results show that the transient phase during the development of the autocatalytic steady-state conduit is much more complex than in non-reacting laminar plumes. Previous experimental work~\cite{ROGpre} has shown that the conduit is conical, rather than cylindrical for the case of non-reacting laminar plumes. This shape is also clear in the simulated autocatalytic plume conduits. The most dramatic feature is of course the pinch-off process. In previous work, it was presumed that pinch-off occurs because of the acceleration of the autocatalytic plume head, which then leads to a ``bottleneck'' in the conduit at the site of pinch-off, resulting in the formation of a second generation head~\cite{ROGprl}. Our simulation results reveal a much more complete picture of the flow development involved in this process. 

As an autocatalytic plume evolves and the conduit elongates, reaction at the interface between reacted and unreacted fluids causes the buoyancy flux to be distributed along the entire length of the conduit. The time-dependent flow in the conduit is shown in Fig.~\ref{spacetime}, which shows a space-time diagram of the vertical velocity profile along the axis of symmetry of the plume. In the early stages of plume growth, before about $23\tau$, the vertical velocity in the conduit increases with distance from the base of the plume. This shows that, even in the early stages, an autocatalytic plume conduit does not play the same role as a non-reacting plume conduit, which delivers buoyancy flux to the more slowly ascending non-reacting plume head. The ascent velocity of the autocatalytic plume head in the early regime exceeds the vertical velocity in the conduit, so that upwelling flow in the conduit does not reach the head. Instead of being fed buoyant fluid by the conduit, an autocatalytic plume head is able to generate its own buoyancy flux by means of reaction with the surrounding fluid.

The later stages of autocatalytic plume development, leading to pinch-off, are more complex. Beyond approximately $23\tau$, a velocity minimum becomes apparent beneath the head. In Fig.~\ref{spacetime}, the minimum is seen as a streak that trails below the maximum near the crest of the plume. The development of the velocity minimum occurs long before morphological indicators of pinch-off are observed, namely, the changes in the growth rates of either $w_h/\ell_h$ or $R_c$, or the visible narrowing of the conduit beneath the head. All of these indicators suggest that pinch-off occurs from anywhere between approximately  $37\tau$ to $41\tau$.  The first generation head continues to accelerate upward, despite the development of the velocity minimum, highlighting the lack of influence that the conduit has on the subsequent evolution of the first generation autocatalytic plume head. The second generation plume head is created by the accumulation of buoyant fluid rising up into the area of the velocity minimum. After it is formed, the second head also accelerates away from the conduit, and a new velocity minimum develops in its wake, bringing the process of head generation back to the beginning of the cycle.  The simulation clearly shows the precursor structure in the conduit velocity that long precedes the morphological indicators of pinch-off. Although it superficially resembles the conduit of a non-reacting plume, the structure below the head of an autocatalytic plume, which might better be called the ``tail", does not contribute in the same way to plume head motion. This is already the case long before pinch-off becomes apparent.

\section{Conclusions}

We have described an experimental and numerical study of buoyant laminar plumes driven by the autocatalytic iodate-arsenous acid reaction. Experimentally, autocatalytic plumes were created in solutions where the viscosity was varied by the addition of glycerol. In the most viscous solutions, pinch-off of the plume head from the conduit was rarely observed, while in the lowest viscosity solutions, pinch-off was not observed. In the mid-range of glycerol concentrations used, however, pinch-off occurred multiple times from a single initiation event. The successive heads all tended to achieve the same upward velocity. The morphology of autocatalytic plume heads was quantified in terms of their radius of curvature, $R_c$ and their width to length aspect ratio, $w_h/\ell_h$. We found that this ratio evolved with time quite differently from that of non-reacting forced plumes. A steep increase in the rate of change of $R_c$ was found to occur at the approximate time when the plume head visibly separates from the conduit. 

To further examine the pinch-off process, axisymmetric simulations of autocatalytic plumes were implemented for the fluid parameters corresponding to a reactant solution with 40\% glycerol. Similar results were shown for both simulated and experimental plumes -- first generation plume heads were observed to accelerate and pinch-off from the conduit, and the metrics used to quantify the morphology of the first generation plume head were in good qualitative agreement with the ones measured for the experimental system. More importantly, the simulation results provided new insight into the pinch-off process. It was found that subtle changes in the velocity structure in the conduit occur long before the appearance of external morphological changes associated with pinch-off. Furthermore, it was determined that the vertical component of the fluid velocity in the conduit is less than the ascent velocity of the head at all times, so that the conduit does not contribute buoyancy to the head but rather forms a tail in its wake. The appearance of velocity minima in this tail are early precursors to the eventual pinch-off of the head. This velocity profile reveals that autocatalytic plume heads are nearly independent flow structures that are driven mainly by the chemical reaction between fluid in the heads and their surroundings.

\begin{acknowledgments}
We thank Anne De Wit for formulating the nondimensionalization scheme used for the simulation, and for valuable discussions on this work. We also thank Matthew Wells, Paul Kushner, Michael Menzinger, Ted Shepherd, and Andrew Belmonte for their useful comments. This research was supported by the Natural Science and Engineering Research Council (NSERC) of Canada.
\end{acknowledgments}

\end{document}